# Record-High Electron Mobility and Controlled Low $10^{15}$ cm$^{-3}$ Si-doping in (010) *β*-Ga$_2$O$_3$ Epitaxial Drift Layers

Carl Peterson,[1,a)] Arkka Bhattacharyya,[1] Kittamet Chanchaiworawit,[1] Rachel Kahler,[1] Saurav Roy,[1] Yizheng Liu,[1] Steve Rebollo,[1] Anna Kallistova,[1] Thomas E. Mates,[1] and Sriram Krishnamoorthy[1,a)]

[1]*Materials Department, University of California Santa Barbara, Santa Barbara, California, 93106, USA*

___________________________

[a)] Author to whom correspondence should be addressed. Electronic mail: carlpeterson@ucsb.edu and sriramkrishnamoorthy@ucsb.edu

We report on metalorganic chemical vapor deposition (MOCVD) growth of controllably Si-doped 4.5 $\mu$m thick *β*-Ga$_2$O$_3$ films with electron concentrations in the $10^{15}$ cm$^{-3}$ range and record-high room temperature Hall electron mobilities of up to 200 cm$^2$/V.s, reaching the predicted theoretical maximum room temperature mobility value for *β*-Ga$_2$O$_3$. Growth of the homoepitaxial films was performed on Fe-doped (010) *β*-Ga$_2$O$_3$ substrates at a growth rate of 1.9 $\mu$m/hr using TEGa as the Gallium precursor. To probe the background electron concentration, an unintentionally doped film was grown with a Hall concentration of 3.43 x $10^{15}$ cm$^{-3}$ and Hall mobility of 196 cm$^2$/V.s. Growth of intentionally Si-Doped films was accomplished by fixing all growth conditions and varying only the silane flow, with controllable Hall electron concentrations ranging from 4.38 x $10^{15}$ cm$^{-3}$ to 8.30 x $10^{15}$ cm$^{-3}$ and exceptional Hall mobilities ranging from 194 – 200 cm$^2$/V.s demonstrated. C-V measurements showed a flat charge profile with the $N_D^+$ - $N_A^-$ values correlating well with the Hall-measured electron concentration in the films. SIMS measurements showed the silicon atomic concentration matched the Hall electron concentration with Carbon and Hydrogen below detection limit in the films. The Hall, C-V, and SIMS data indicate the growth of high-quality 4.5 $\mu$m thick *β*-Ga$_2$O$_3$ films and controllable doping into the mid $10^{15}$ cm$^{-3}$ range. These results demonstrate MOCVD growth of electronics grade record-high mobility, low carrier density, and thick *β*-Ga$_2$O$_3$ drift layers for next generation vertical *β*-Ga$_2$O$_3$ power devices.

As global power consumption continues to rise[1], solutions for creating highly efficient power electronic devices become critical to reduce the amount of energy wasted during power conversion. For decades, semiconducting materials such as GaN and SiC have demonstrated the inherent advantages of having a wide bandgap (WBG) when making more efficient power devices[2]. A wider bandgap allows a material to sustain higher electric fields, and thus, devices can be made smaller and less resistive. Ultra-Wide Band Gap semiconductors (UWBG) have shown promise to continue to push the boundaries of power device efficiency[3]. A contender for the next-generation UWBG material is Beta Gallium Oxide ($\beta$-$Ga_2O_3$), with a bandgap of 4.6-4.9 eV and a predicted critical electric field strength of 8 MV/cm. Figure of Merit analyses of WBG and UWBG materials show $\beta$-$Ga_2O_3$ to be the best-in-class material[4]. $\beta$-$Ga_2O_3$ is also the only WBG/UWBG material to demonstrate melt-grown conductive and insulating bulk substrates with dopant impurity control[5–11], enabling large area and low extended defect density substrate platforms, reduced epitaxial defect density, and potentially cheaper production costs.

An epitaxial growth technique that has shown great promise and versatility for $\beta$-$Ga_2O_3$ growth is metalorganic chemical vapor deposition (MOCVD) due to its scalable growth rates[12–21], high electron mobilities[12,14,22–26], material alloying[27,28], in-situ etching[29], in-situ dielectrics[30,31], wide range of n-type conductivity[32,33], and delta doping capabilities[34,35]. Power devices fabricated using MOCVD epitaxial films have also been demonstrated to have state-of-the-art performance[36–46]. One major requirement for creating vertical power devices rated to sustain 10s of kilovolts is the ability to grow tens of microns of high-quality epitaxial material with very low background impurity concentration, very low compensating species concentration, controlled low doping (~$10^{15}$ -$10^{16}$ cm$^{-3}$), and high carrier mobility. MOCVD $\beta$-$Ga_2O_3$ has recently shown the capability and promise to meet these requirements[12,13,16,17,19], with minimally compensated multi-$\mu$m thick unintentionally doped (UID) epilayers with electron concentrations as low as 2.4 x $10^{15}$ cm$^{-3}$ having already been demonstrated[12]. In this work, we demonstrate controllable silicon doping with electron concentrations as low as 4.38 x $10^{15}$ cm$^{-3}$ in 4.5 $\mu$m thick $\beta$-$Ga_2O_3$ epilayers with record-high room temperature Hall mobilities reaching up to 200 cm$^2$/V.s.

Growth of the $\beta$-$Ga_2O_3$ epitaxial films was done using an Agilis 100 cold wall MOCVD reactor from Agnitron Technology Inc. The vertical quartz wall reactor was equipped with a remote injection showerhead (RIS) with a showerhead to susceptor distance of ~18 cm. The gallium precursor used for growth was TEGa, ultra-high purity $O_2$ gas (5N) was chosen as the oxygen source, dilute silane was ($SiH_4$) used as the silicon source, and high purity Argon (5N) was chosen as the carrier gas. The homoepitaxial growths were performed on 5x5 mm$^2$ Fe-doped semi-insulating (010) $\beta$-$Ga_2O_3$ substrates commercially acquired from Novel Crystal Technology Inc., Japan. Prior to growth, substrates were cleaned using solvents (Acetone, Methanol) followed by a de-ionized water rinse. After solvent cleaning, the substrates were submerged in a 49% HF solution for ~30 minutes[22] and then loaded into the MOCVD chamber for growth within 10 minutes of acid cleaning to prevent Si contamination at the growth interface[47]. After samples were loaded into the MOCVD, a 10-minute pre-growth anneal was performed at 950 ºC in an $O_2$/Ar gas ambient as an additional growth interface treatment step. Growth on all samples was performed in the mass-transport limited regime at a temperature of 900 ºC with a growth chamber pressure of 15 Torr. Additionally, for all samples the $O_2$, TEGa, and Ar flow rates were fixed at 1000 sccm, 95.58 $\mu$mol/min, and 1500 sccm respectively, leading to a VI/III ratio of 428. These reactor conditions led to a growth rate of 1.9 $\mu$m/hr. Growth rate was verified by loading a c-plane sapphire substrate into the chamber along with the $\beta$-Ga$_2$O$_3$ (010) substrates and performing cross-sectional SEM imaging on the $\beta$-Ga$_2$O$_3$ on sapphire sample to determine the epilayer thickness. See supplementary information figure S1 for a cross-sectional SEM image.

After growth, isolated Van Der Pauw structures were fabricated on all samples. An etch mask was created using 700 nm of PECVD SiO$_2$ followed by a lithographically defined 10/120 nm Ti/Ni metal stack deposited via electron-beam (E-Beam) evaporation. After metal lift-off, the SiO$_2$ was etched using a CHF$_3$ inductively coupled plasma (ICP) dry etch. The $\beta$-Ga$_2$O$_3$ epitaxial films were etched via a BCl$_3$ ICP etch with the etch continuing into the Fe-doped substrate for effective isolation. The etch depth was confirmed using profilometry as 5.2 $\mu$m. Finally, lithographically defined Ti/Au (20/150 nm) Ohmic contacts were deposited via E-Beam and a 1 minute 470 ºC Anneal was performed in N$_2$ ambient to improve the ohmic contacts. Pre-and post-anneal I-V data are shown in supplementary information Figure S2.

Since reactor conditions such as the gas flow rates and reactor pressure have been shown to effect the background electron concentration in low-doped $\beta$-Ga$_2$O$_3$ epitaxial films[12], the first sample grown, Sample A, was grown with zero silane flow to determine the unintentionally doped (UID) background electron concentration for the given growth conditions. Hall measurements on the 4.8 $\mu$m thick UID sample A were conducted via a Lake Shore Hall measurement system. The measured Hall carrier concentration and mobility for sample A were 3.43 x 10$^{15}$ cm$^{-3}$ and 196 cm$^2$/V.s respectively. After determining the background UID electron concentration for the given growth conditions, four 4.5 $\mu$m thick intentionally Si doped samples (B-E) were grown. For the intentionally doped samples, all growth conditions were fixed (temperature, pressure, TEGa/Ar/O$_2$ gas flow rates) except for the silane flow which was varied from 6.55 – 34.5 pmol/min. The intentionally doped films had Hall electron concentrations ranging from 4.38 x 10$^{15}$ cm$^{-3}$ to 8.30 x 10$^{15}$ cm$^{-3}$ and record-high room temperature Hall mobilities ranging from 194 – 200 cm$^2$/V.s. The numerical values of the measured Hall data and silane flow rates for samples A-E are given in Table I. Figure 1(a) shows the linear relationship between the Hall carrier density vs. silane flow for the intentionally doped samples B-E where the UID condition, sample A, is represented by a dotted red line. The dashed green line with a slope of one in Figure 1(a) shows the linear dependence of carrier density and silane flow. A linear fit with a slope of one (in a log-log plot) is expected because Si incorporation into $\beta$-Ga$_2$O$_3$ has been experimentally observed to follow the Langmuir adsorption model, where the Si adatoms and Ga adatoms compete for group III sites during growth[48,49]. Thus, if the concentration of intentional Si-doping is greater than the background electron concentration, the expected Si doping (and thus, electron concentration) should depend only on the ratio of Si/Ga precursors, leading to an expected 1:1 linear relationship between silane flow and electron concentration (since TEGa flow is fixed in this experiment). The data points in Figure 1(a) generally follow this trend, with the outlier, sample B, most likely not following the trend because it is too close to the UID background. Figure 1(b) represents the measured Hall mobility vs. Hall carrier density for all samples and shows that the mobility was nearly constant across all electron concentration/silane flow values. All measured Hall mobility values were close or equal to the theoretical maximum value for phonon scattering-limited mobility in $\beta$-Ga$_2$O$_3$ (~200 cm$^2$/V.s)[50]. The current-voltage (I-V) characteristics of the annealed Hall pads in the inset Figure 1(c) were measured via a Keithley 4200 parameter analyzer and showed highly linear ohmic behavior.

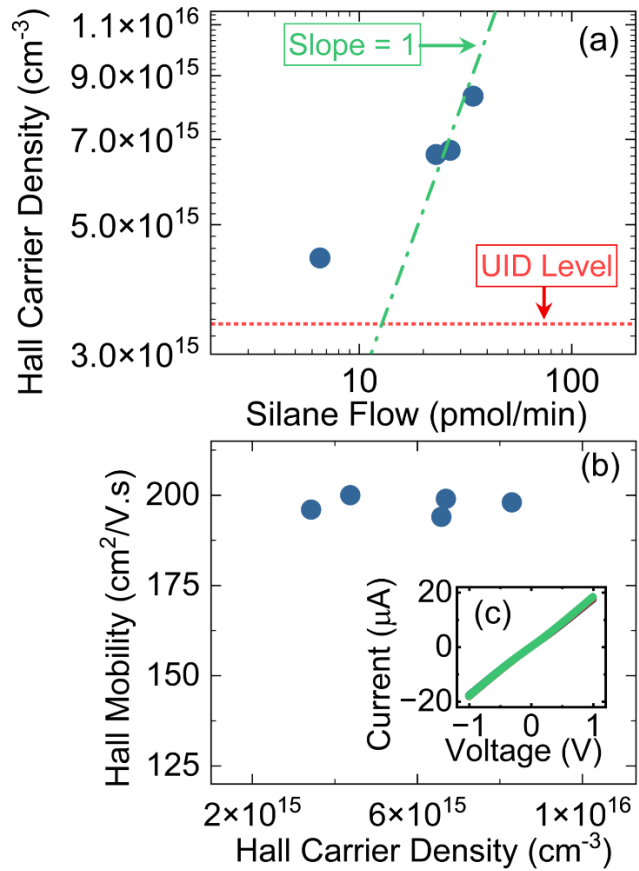

FIG. 1. (a) Log-log plot of Hall carrier density vs. silane flow for the intentionally doped samples, where the green dashed line corresponds to the expected doping profile, or a slope equal to one, and the red dotted line corresponds to the carrier density of the UID sample with no silane flow. (b) Plot of the Hall mobility vs Hall carrier density for the intentionally and unintentionally doped samples. The inset plot (c) is the current-voltage plot of the Hall pads, showing linear ohmic contacts after annealing.

The surface morphology of the thick epilayers was characterized using an Asylum Research atomic force microscope (AFM), with a representative 5 x 5 $\mu$m scan of sample B depicted in Figure 2(a). RMS surface roughness values of 5.12 – 7.55 nm were observed across the 4.5 $\mu$m thick films. The surface scans showed clear step bunching behavior along the [001] crystal orientation which is consistent with other MOCVD-grown and MBE-grown films grown on (010) oriented $\beta$-$Ga_2O_3$ substrates[12,16,17,20,22,51,52]. Figure 1(b) shows a large area optical microscope image of sample B using differential interference contrast (DIC) microscopy to better resolve the defects present on the surface. These defects are hillocks oriented in the [001] crystal orientation, similar to those seen in literature[20] and are potentially due to adduct formation and contamination during growth, highlighting the importance of cleaning the MOCVD chamber walls, showerhead, and susceptor between thick $\beta$-$Ga_2O_3$ growths using TEGa. Further analysis of the defects is discussed in the supplementary information, with Figure S3 showing the AFM scans and microscope images for all samples and Figure S4 showing an AFM scan of a single defect. To mitigate the formation of adducts and particulate contamination during growth one can reduce the growth pressure further or decrease the distance between the showerhead and the susceptor to a close-coupled showerhead (CCS) system to minimize gas phase pre-reactions. Ta-Shun Chou et al. have investigated the latter method and showed that reducing the showerhead distance from 8 to 1.5 cm significantly reduced particulate contamination and doubled the growth rate, highlighting the advantage of a CCS system over an RIS system for particulate-free thick growths[53]. However, despite the surface having some particulate contamination, the RIS system was able to accomplish 4.5 $\mu$m thick growths with low electron concentrations and exceptionally high mobilities, highlighting the capabilities of using an RIS system to grow drift layers.

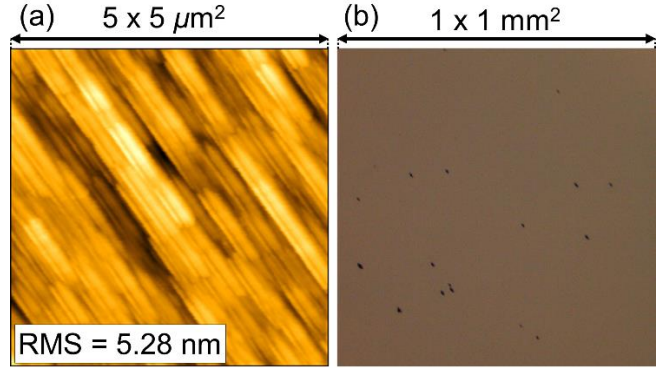

FIG. 2. Representative (a) AFM scan and (b) DIC optical microscope image for sample B.

TABLE I. Summary of room temperature electronic transport characterization

| Sample Name | Sample Thickness ($\mu$m) | Silane Flow (pmol/min) | Hall Carrier Concentration (cm$^{-3}$) | Hall Mobility (cm$^2$/V.s) | Charge Density from CV measurements $N_D^+ - N_A^-$ (cm$^{-3}$) |
|---|---|---|---|---|---|
| A | 4.8 | 0 | 3.43 x 10$^{15}$ | 196 | 3.00 x 10$^{15}$ |
| B | 4.5 | 6.55 | 4.38 x 10$^{15}$ | 200 | 4.45 x 10$^{15}$ |
| C | 4.5 | 23.1 | 6.59 x 10$^{15}$ | 194 | 6.70 x 10$^{15}$ |
| D | 4.5 | 26.9 | 6.70 x 10$^{15}$ | 199 | 7.80 x 10$^{15}$ |
| E | 4.5 | 34.5 | 8.30 x 10$^{15}$ | 198 | 1.00 x 10$^{16}$ |

Large area (0.164 mm$^2$) lateral Schottky C-V pads were lithographically defined, and a Schottky metal stack of 50/150 nm of Ni/Au was deposited via E-beam evaporation. The apparent charge density ($N_D^+ - N_A^-$) values in Table I and Figure 3(a) were extracted by averaging the C-V charge profile found in Figure 3 (b) and were plotted against silane flow. The values correlate well with the values of Hall carrier density shown in Table I and Figure 1(a) as well as followed the expected linear relationship between silane flow and charge concentration. The apparent C-V charge profile in Figure 1(b) was flat, showing a nearly constant $N_D^+ - N_A^-$ value at all depth values.

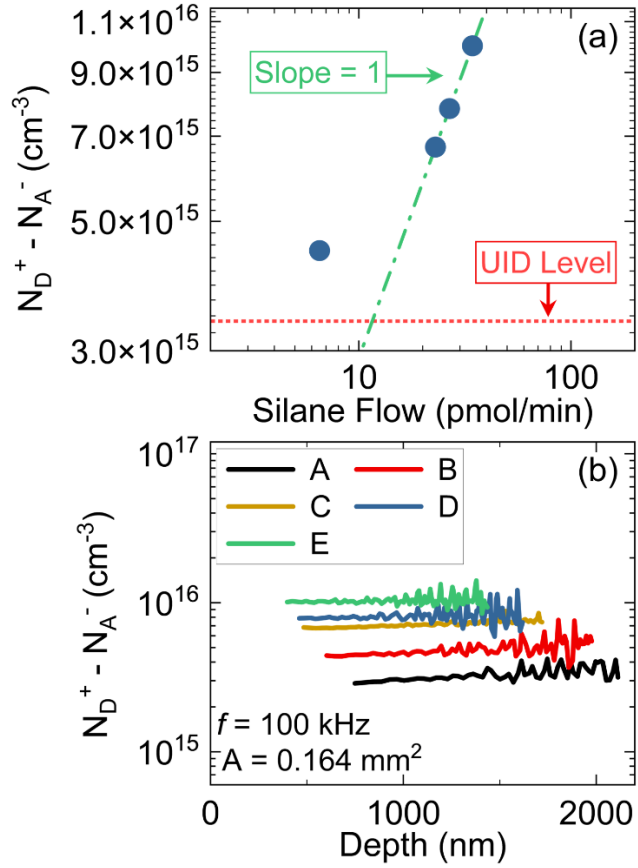

FIG. 3. (a) Averaged $N_D^+ - N_A^-$ vs. silane flow for the intentionally doped samples, where the green dashed line corresponds to the expected doping profile, or a slope equal to one, and the red dotted line corresponds to the carrier density of the UID sample with a silane flow of zero. (b) Apparent C-V charge profile for all samples measured at 100 kHz, showing a flat charge profile throughout the film.

To better understand the impurity concentrations in the low-doped and high mobility epilayers, secondary ion mass spectroscopy (SIMS) was performed using a CAMECA IMS 7F tool and $Cs^+$ primary beam ions. SIMS was done on a 4.8 $\mu$m UID calibration sample with the only growth variable differences being an $O_2$ flow rate of 1200 sccm and a growth temperature of 950 °C. The measured Hall electron concentration for this sample was 3.96 x $10^{15}$ $cm^{-3}$. Figure 4 indicates an atomic Si concentration of ~ 3-4 x $10^{15}$ $cm^{-3}$ measured from SIMS which closely matches the measured electron concentration from Hall results. The corroboration between the measured electron concentration and atomic silicon concentration indicates that there is little to no compensation of donors or additional donor impurities other than Si in the films and there is full activation of silicon atoms in the grown films[54]. The amount of Carbon and Hydrogen impurities in the films was measured to be at the detection limit of the SIMS tool, as shown in the supplementary information Figure S5. Thus, we can speculate that the presence/effect of compensating species or impurities is negligible, as the measured Hall mobilities in the epilayers are close to the theoretical maximum value for $Ga_2O_3$, confirming the growth of high-purity material. Nevertheless, detailed temperature-dependent Hall characterization and carrier statistics analysis is required in the future to quantify compensation in these films.

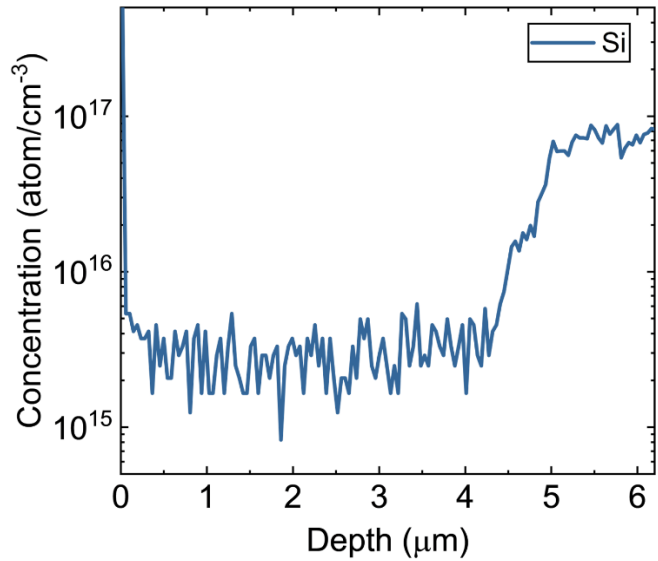

FIG. 4. SIMS profile showing the atomic concentration of silicon for a 4.8 μm thick UID sample. The measured Si concentration matches well with the Hall carrier density (3.96 x $10^{15}$ cm$^{-3}$).

Figure 5 shows the room temperature Hall mobility vs. Hall carrier density benchmarking for state-of-the-art β-Ga$_2$O$_3$ homoepitaxial films grown by a variety of different methods. Results indicate that the mobility values measured in this study are highest reported in β-Ga$_2$O$_3$ at any doping level, which again suggests the films grown in this work have very low compensation.

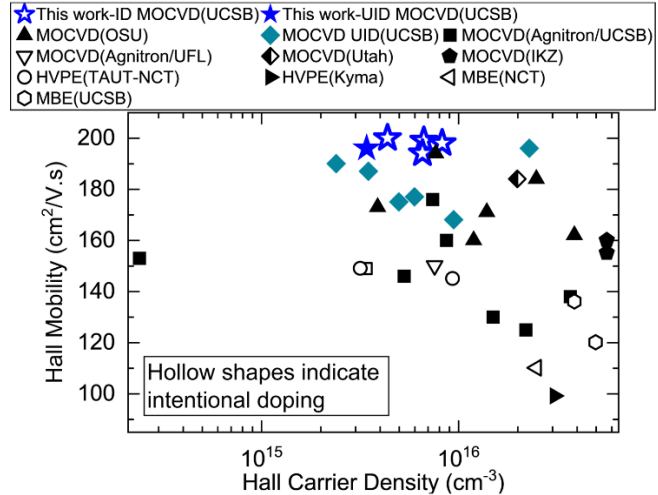

FIG. 5. Benchmarking plot of measured room temperature Hall mobility vs. Hall carrier density with state-of-the-art low-doped results from MOCVD (UCSB[12,22], OSU[25,26], Agnitron/UCSB[14,18,24,32], Agnitron/UFL[19], University of Utah[23], IKZ[13,53]), HVPE (TAUT-NCT[55], Kyma[56]), and MBE (NCT[57], UCSB[58,59]). Hollow shapes indicate intentional doping (ID) and solid shapes indicate UID films.

In conclusion, we have demonstrated the ability to grow 4.5 μm thick high-quality MOCVD β-Ga$_2$O$_3$ epitaxial films with controllable doping into the $10^{15}$ cm$^{-3}$ range with record-high mobilities up to 200 cm$^2$/V.s. A 4.8 μm thick UID film determined the background Hall electron concentration to be 3.43 x $10^{15}$ cm$^{-3}$ and had a Hall mobility of 196 cm$^2$/V.s. Four 4.5 μm thick intentionally Si doped samples (B-E) were grown with the silane flow varying from 6.55 – 34.5 pmol/min and the rest of the growth parameters fixed, leading to Hall electron concentrations ranging from 4.38 x $10^{15}$ cm$^{-3}$ to 8.30 x $10^{15}$ cm$^-$ and Hall mobilities ranging from 194 – 200 cm$^2$/V.s. AFM scans showed a surface roughness ranging from 5.12 – 7.55 nm with optical microscopes showing some particulate contamination. The combination of a flat C-V doping profile showing 10$^{15}$ cm$^{-3}$ electron concentration, C and H concentrations below the detection limit in SIMS, and Hall mobility values reaching the predicted theoretical maximum suggests the growth of high-quality *β*-Ga$_2$O$_3$ films. This work represents a significant milestone in advancing MOCVD *β*-Ga$_2$O$_3$ epitaxy toward use in device drift layers for high voltage rated power switches.

**SUPPLEMENTARY MATERIAL**

The supplementary material contains additional data and discussions regarding the cross-sectional SEM image of a growth rate calibration sample, annealed vs. un-annealed Hall pad I-V profiles, four-corner vs. mesa isolated Van Der Pauw Hall structures, AFM and DIC microscope surface analysis of all samples, and SIMS measurements of background impurities at detection limit.

**ACKNOWLEDGMENTS**

The authors acknowledge funding from the ARPA-E ULTRAFAST program (DE-AR0001824) and Coherent / II-VI Foundation Block Gift Program. A portion of this work was performed at the UCSB Nanofabrication Facility, an open access laboratory. Carl Peterson would like to acknowledge Ziliang Ling for assistance with measurements.

**DATA AVAILABILITY**

The data that support the findings of this study are available from the corresponding author upon reasonable request.

# Supplementary Material

# Record-High Electron Mobility and Controlled Low $10^{15}$ cm$^{-3}$ Si-doping in (010) *β*-Ga$_2$O$_3$ Epitaxial Drift Layers

Carl Peterson,[1,a)] Arkka Bhattacharyya,[1] Kittamet Chanchaiworawit,[1] Rachel Kahler,[1] Saurav Roy,[1] Yizheng Liu,[1] Steve Rebollo,[1] Anna Kallistova,[1] Thomas E. Mates,[1] and Sriram Krishnamoorthy[1,a)]

[1]*Materials Department, University of California Santa Barbara, Santa Barbara, California, 93106, USA*

___________________________

[b)] Author to whom correspondence should be addressed. Electronic mail: carlpeterson@ucsb.edu and sriramkrishnamoorthy@ucsb.edu

**S1. Cross-Sectional SEM Thickness Information**

Cross sectional SEM images were taken to calibrate the growth rate of the thick *β*-Ga$_2$O$_3$ epitaxial layers. Figure S1 shows the cross section of a c-plane sapphire sample co-loaded with the *β*-Ga$_2$O$_3$ Fe-doped substrates, clearly displaying the epilayer thickness.

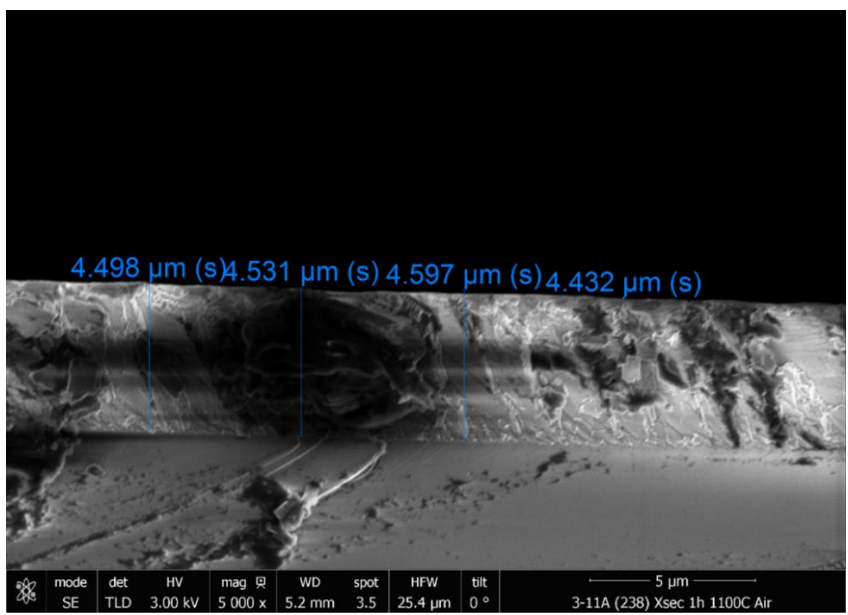

*Fig. S1. Cross sectional SEM of β-Ga$_2$O$_3$ on sapphire, co-loaded with the thick growths to calibrate growth rate and film thickness.*

**S2. Hall I-V Annealing Data & 4-Corner Vs. Isolated Hall**

The hall pads were annealed in an N$_2$ ambient for 1 min at 470 °C in order to improve the ohmic contacts to the low electron concentration films. Figure S2 shows the major improvement of the linearity of the ohmic contact, highlighting the importance of annealing to make good contact to low-doped $\beta$-$Ga_2O_3$ material.

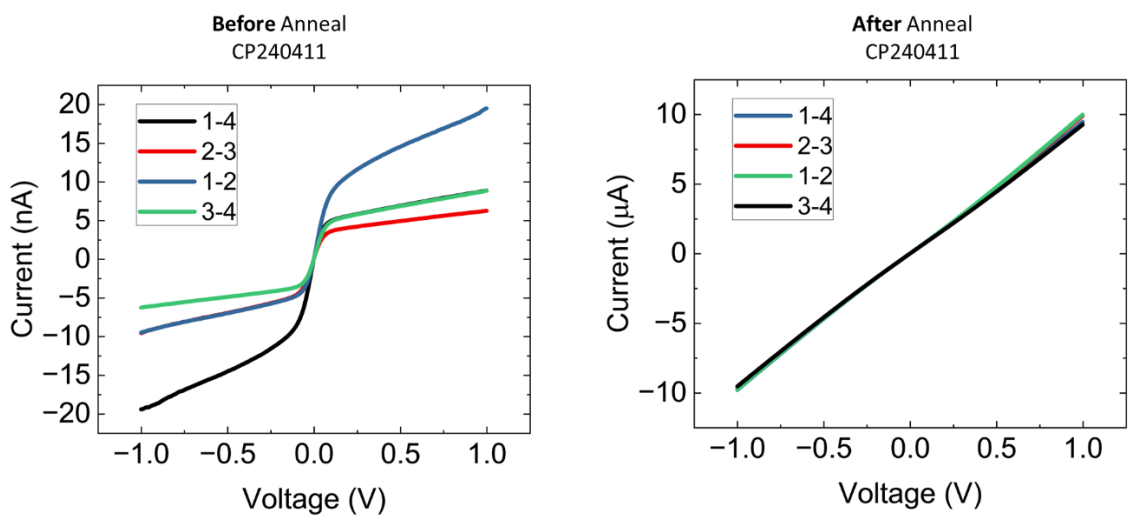

*Fig. S2. I-V measurements on hall pads before and after annealing*

## S3. Four-Corner Vs. Mesa Isolated Van Der Pauw Hall

For thick films, lithographically defined isolated Van Der Pauw test structures were found to provide the most accurate Hall measurements, compared to four-corner Hall measurements. Details of Hall measurements are presented below in Table S1.

TABLE S1. Summary of room temperature electronic transport characterization, comparing 4-corner, isolated but non-annealed, and isolated plus annealed Van Der Pauw Hall structures

| Sample Name | Silane Flow (pmol/min) | 4-Corner Hall Carrier Concentration ($cm^{-3}$) | 4-Corner Hall Mobility ($cm^2/V.s$) | Non-Annealed Isolated Hall Carrier Concentration ($cm^{-3}$) | Non-Annealed Isolated Hall Mobility ($cm^2/V.s$) | Annealed + Isolated Hall Carrier Concentration ($cm^{-3}$) | Annealed + Isolated Hall Mobility ($cm^2/V.s$) |
|---|---|---|---|---|---|---|---|
| A | 0 | $4.47 \times 10^{15}$ | 164 | $3.40 \times 10^{15}$ | 196 | $3.43 \times 10^{15}$ | 196 |
| B | 6.55 | $5.13 \times 10^{15}$ | 174 | $4.31 \times 10^{15}$ | 199 | $4.38 \times 10^{15}$ | 200 |
| C | 23.1 | $7.99 \times 10^{15}$ | 165 | $6.55 \times 10^{15}$ | 196 | $6.59 \times 10^{15}$ | 194 |
| D | 26.9 | $7.30 \times 10^{15}$ | 184 | $7.08 \times 10^{15}$ | 189 | $6.70 \times 10^{15}$ | 199 |
| E | 34.5 | $9.29 \times 10^{15}$ | 177 | $8.23 \times 10^{15}$ | 198 | $8.30 \times 10^{15}$ | 198 |

## S4. AFM and Surface Defect Morphology

AFM scans were performed on all samples to obtain the surface morphology and RMS surface roughness values. Figure S3 displays 5x5 µm AFM surface scans that showed clear step bunching and RMS roughness values ranging from 5.12 – 7.55 nm for the 4.5 µm thick films. Figure S3 also shows DIC optical microscope images of the sample surface. There is a general trend observed where an increased number of particles were present on the surface the more samples that were grown (see date of sample growth arrow). This implies that there was a buildup of adducts in the reactor (showerhead, reactor walls, susceptor) that were not removed from general reactor cleaning procedures which affect future growths.

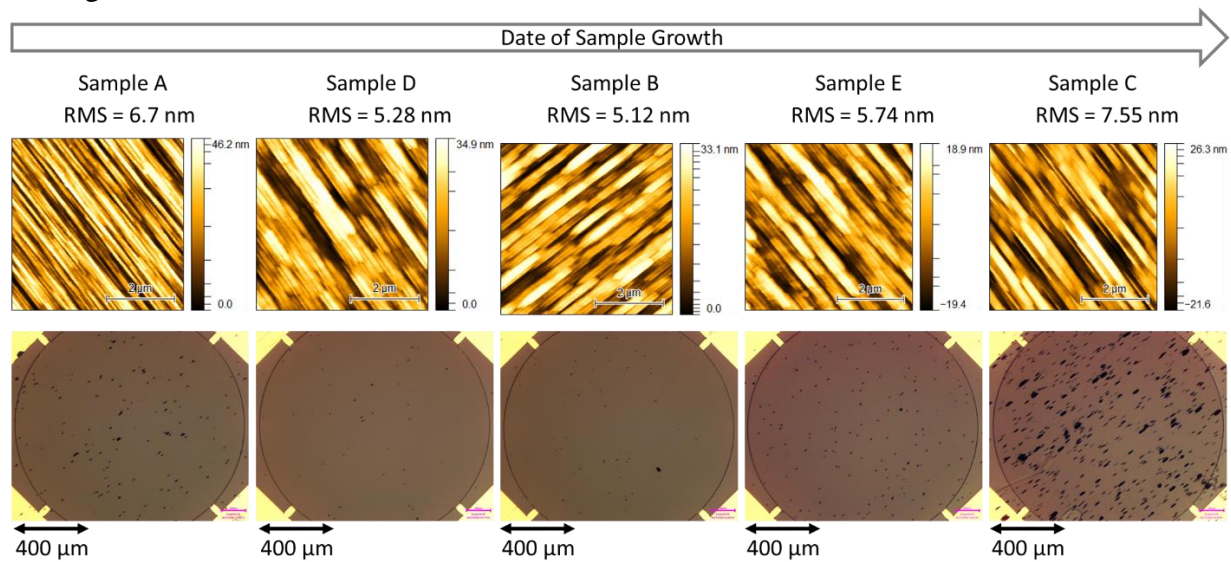

*Fig. S3. Surface AFM scans and DIC optical microscope images of the sample surface.*

Figure S4 shows the 3D structure of the surface defects using AFM scanning. The dark defects seen in figure S3 appear to be ridges aligning with the step bunching of the surface morphology. These ridges appear to arise from the deposition of a particulate on the surface during growth, which interrupts the growth mode.

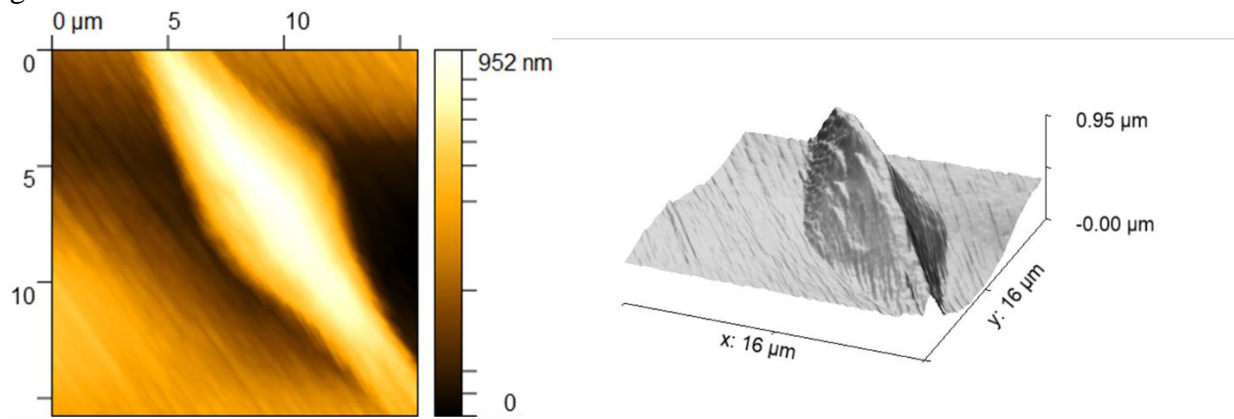

*Fig. S4. AFM scan and 3D map of a single defect*

## S5. SIMS Scan for Carbon and Hydrogen

A UID epitaxial film was deposited via the standard growth parameters described in the main manuscript on a Sn-Doped $\beta$-$Ga_2O_3$ (010) substrate. A Sn-doped substrate was chosen to improve the conductivity of the sample for SIMS measurement to prevent charging effects. SIMS was performed using a CAMECA IMS 7F tool and the carbon and hydrogen were found to be at the limits of the tool detection limit, as seen in Figure S5.

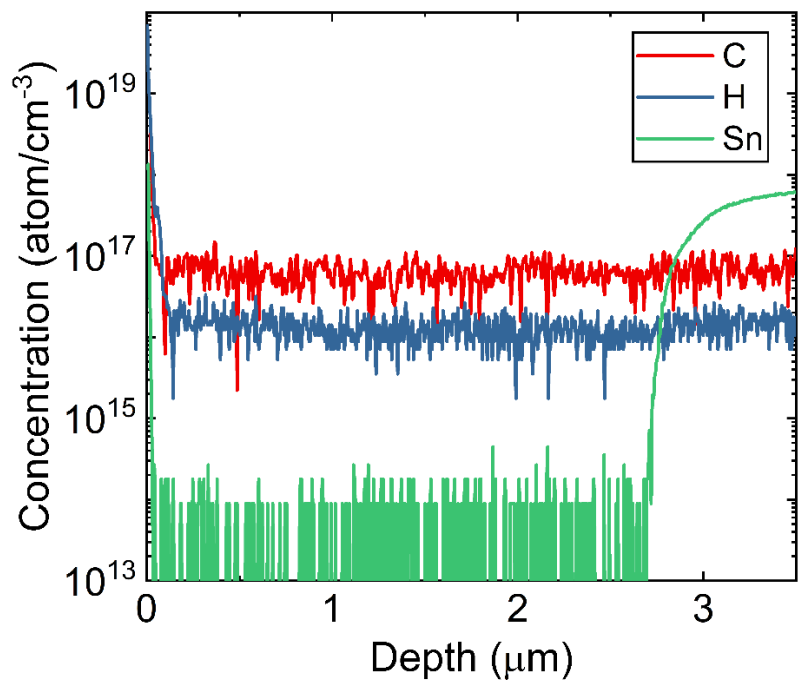

*Fig. S5. SIMS measurement showing C and H impurities at the tool detection limit*